\documentclass[aps,twocolumn,superscriptaddress,showpacs]{revtex4}
\usepackage{epsfig}
\usepackage{changebar}
\usepackage{graphicx}
\usepackage{epsfig}
\usepackage{subfigure}
\usepackage{amsmath}
\usepackage{rotating}
\usepackage{amsfonts}
\usepackage{amssymb} 
\newlength {\oldtextheight}
\newlength {\oldheadsep}
\psfigdriver{dvips}

\begin{document}
 
\begin{widetext}
 
\begin{center}
{\Large Self-similar correlation function in brain resting-state fMRI}\\

\vspace{1cm}
Paul Expert$^{1,2}$, Renaud Lambiotte$^{1}$, Dante R. Chialvo$^{3}$, Kim Christensen$^{1,2}$, Henrik Jeldtoft Jensen$^{1,4,*}$\\
David J. Sharp$^{5}$ and  Federico Turkheimer$^{5}$\\
\vspace{1cm}
(1){Institute for Mathematical Sciences, 53 Prince's Gate, Exhibition Road\\ Imperial College London, London SW7 2PG, UK}\\
(2){Blackett Laboratory, Prince Consort Road, Imperial College London, London SW7 2AZ, UK}\\
(3) {Department of Physiology, Northwestern University, Chicago, Illinois 60611, USA}\\
(4) {Department of Mathematics, Queen's Gate, Imperial College London, London SW7 2AZ, UK}\\
(5){Centre for Neuroscience, Department of Experimental Medicine and Toxicology, Hammersmith Campus\\ DuCane Road, Imperial College London, London W12 0NN, UK} \\
\vspace{.5cm}
(*)To whom correspondence should be addressed. E-mail: h.jensen@imperial.ac.uk
\vspace{1cm}
\end{center} 
\begin{center} {\Large Abstract} \end{center}
{\large Adaptive behavior, cognition and emotion are the result of a be\-wil\-der\-ing variety of brain spatiotemporal activity patterns.  
An important problem in neuroscience is to understand the mechanism by which the human brain's 100 billion neurons 
and 100 trillion synapses manage to produce this large repertoire of cortical configurations in a flexible manner.
In addition, it is recognized that temporal correlations across such configurations cannot be arbitrary,
but they need to meet two conflicting demands: while diverse cortical areas should remain functionally
segregated from each other, they must still perform as a collective, i.e., they are functionally integrated.
Here, we investigate these large-scale dynamical properties by inspecting the character of the spatiotemporal
correlations of brain resting-state activity. In physical systems, these correlations in space and time are
captured by measuring the correlation coefficient between a signal recorded at two different points
in space at two different times. We show that this two-point correlation function extracted
from resting-state fMRI data exhibits self-similarity in space and time.
In space, self-similarity is revealed by considering three successive spatial coarse-graining steps
while in time it is revealed by the $1/f$ frequency behavior of the power spectrum. 
The uncovered dynamical self-similarity implies that the brain is spontaneously at a
continuously changing (in space and time) intermediate state between two extremes,
one of excessive cortical integration and the other of complete segregation.
This dynamical property may be seen as an important marker of brain well-being both in health and disease.}
 
\vspace{1cm}

\end{widetext}
\thispagestyle{empty} 
It is increasingly evident that brain regions are continuously 
interacting even when the brain is ``at rest'' and, more importantly, that the 
functional networks uncovered from resting data closely 
matches those derived from a wide variety of different activation conditions \cite{fox2007,Beckmann-2009}.
Starting with the uncovering of coherent fluctuations of 
functional magnetic resonance imaging (fMRI) in time series of motor 
cortex \cite{biswal}, many other findings  have validated the
notion of correlated networks as a dynamical substrate of the resting brain.
It has been established that these networks, which can be separated on the
basis of their temporal features \cite{xiong,cordes,beckmann2005}, are located at consistent locations across subjects and are equally detectable even during sleep \cite{fuku} and anesthesia \cite{vincent}. 

These exciting findings provide a novel window to observe the brain at work and, at the same time, highlight our limited understanding of the functional organization of the brain at large scales \cite{werner}, compared with the, often precise, knowledge  we have of the small (neural circuit level) scale. 
In particular, little is known on how the cortex is able to solve the conflicting dynamical demands imposed by the functional segregation of local areas differing in their anatomy and physiology on one side and on the other their global integration shown during perception and behavior. This riddle is clearly pointed out by Tononi  \cite{tononi}: ``traditionally, localizationist and holist views of brain function have exclusively emphasized evidence for either functional segregation \emph {or} for functional integration among  components of the nervous system. Neither of these views  alone adequately accounts for the multiple levels at which  interactions occur during brain activity''. 
In connection with this dilemma, it has been suggested  that the brain's conflicting demands are a generic
property of many collectives, regardless of being composed by neurons, genes, individuals, etc \cite{Dante-2007,Dante-2009}.
This implies that the search for the mechanism behind this dilemma must be guided by the general
properties of the system, rather than by the details of the neurobiology. In that regard, the study of large scale
collective properties have a long tradition in statistical physics, allowing the
identification of different dynamical regimes through the study of ubiquitous correlation properties.
In this work, we are guided along that direction, and the paper is dedicated to report the
spatiotemporal correlation properties of fMRI resting-state data which are found to exhibit
robust self-similarity signatures in space and time.
This finding has profound significance, because it demonstrates a continuous range
of correlations between the local cortical circuits up to the entire brain.
Thus, this is the first direct empirical demonstration of the brain resting-state activity
exhibiting \emph{simultaneously} functional segregation \emph {and} integration. 

A variety of experiments have already provided indications of self-similarity in spatial and temporal scales of the brain dynamics.  For example, at small scale, avalanches of neuronal activity in rat \cite{Beggs-2003,Beggs-2004} and monkey \cite{peterman} cortex are known to be scale-free, a finding that has been modeled by Arcangelis {\em et al.} \cite{Arcangelis-2006}, Levina {\em et al.}  \cite{levina} and Buice and Cowan \cite{Buice-2007} among others. At larger scale, it was reported that brain fMRI networks are characterized by scale invariant degree distributions \cite{Dante-2005} as described by Egu{\'i}luz and collaborators and later replicated and extended by van den Heuvel {\em et al.} \cite{vandenHeuvel}. These networks are indistinguishable from those extracted from model systems
at criticality \cite{Dante-2009,Dante-2008}. Other findings include the observation of $1/f$ power spectra from simultaneously recorded magnetoencephalography and electroencephalography signals \cite{Linkenkaer-Hansen-2001}, fMRI signals \cite{Bullmore-2004-2,Bullmore-2005,Kitz} as well as from cognitive responses \cite{VanOrden}, all indicative of long range correlations in brain dynamics processes.
It is these myriads of observations of power-law behavior and long range correlations that have led
to the conjecture that the human brain
as a whole behaves as a system at criticality \cite{Dante-2004,Dante-2006,Bak}
and that the framework of self-organized 
criticality \cite{Bak,Jensen,Christensen} may be relevant to understand large-scale brain dynamics.
In that sense, critical dynamics endows the system with a high
susceptibility and a broad repertoire of responses, natural requirements for healthy brain function.

Here, we analyze directly the correlation function of the voxel signals in space and time.
A voxel is assigned a set of three weights describing the
density of grey matter, white matter and cerebrospinal fluid (CSF) within the voxel.
Up to now, the standard analysis of fMRI functional correlations only included
voxels with a density of grey matter above a certain threshold
value. In the present analysis, we do not discard any voxels, nor do we apply a threshold on the
correlation coefficients. Hence, the present analysis corresponds to the standard analysis tool applied in 
physics and materials science when structural properties are investigated.

\begin{figure}[h]
\begin{center}
\includegraphics[width=0.38\textwidth]{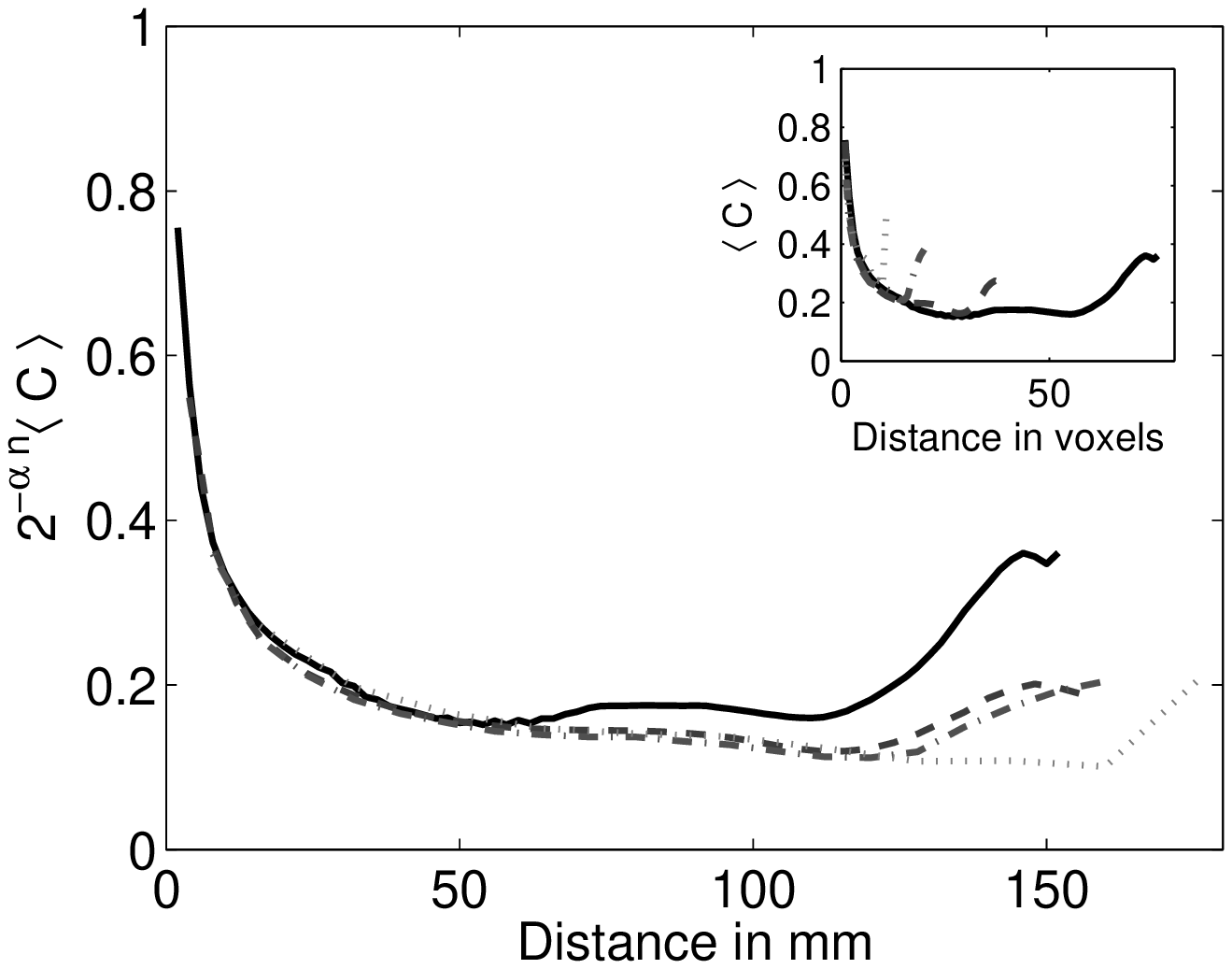} \\
\end{center}
\vspace*{-0.5cm}
\begin{center}
\includegraphics[width=0.38\textwidth]{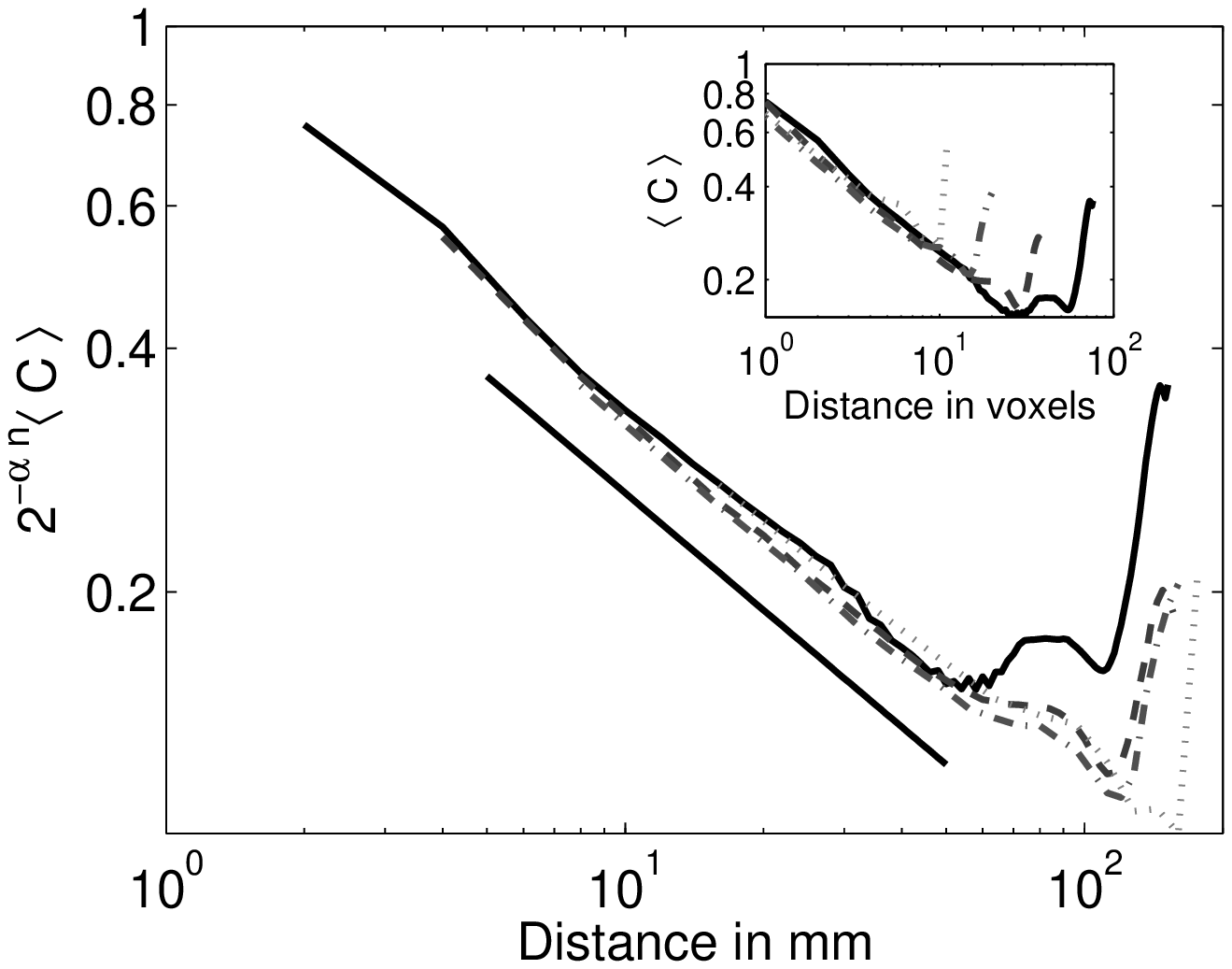}
\end{center}
\caption{The renormalized average correlation function $ 2^{-\alpha n} \langle C \rangle$ vs real distance
(inset: $\langle C \rangle$ vs voxel distance) for the four levels of description for one subject.
Full line: $128\times 128\times 31$ ($n=0$), dashed line: $64\times 64\times 16$ ($n=1$), dashed-dotted
line: $32\times 32\times 8$ ($n=2$) and dotted line: $16\times 16\times 4$ ($n=3$).
Top panel linear-linear and bottom panel log-log axis, respectively. Full straight line is a guide to the eye for a power
law with exponent $\beta=0.47\pm0.2$. Parameter $\alpha = 0.45$.}\label{Fig1}
\end{figure}
To investigate whether the system is self-similar in space, we perform a coarse-graining analysis directly on the voxel data and then
extract the associated correlation function. We will show that 
the coarse-grained correlation function exhibits a power-law decay of correlations at length scales
between the microscopic scale of a voxel and the macroscopic scale of the size
of the brain. This self-similarity is reminiscent
of the multi-scale modular organization \cite{Kaiser-1,Kaiser-2} observed in fMRI data \cite{meunier}, but also
a robust finding in systems near the critical point \cite{Bak,Jensen,Christensen}.

\section{Results}
\emph{Spatial Self-similarity:} Starting from the computation of the correlation matrix, the average correlation function for each
separation $r$ can be estimated. To study the degree of self-similarity present in the correlation function, we
perform a coarse-graining procedure and check for self-similarity via Eq. \eqref{EqScaling} as detailed in the 
section of Material and Methods.
The result of this analysis for a single subject is shown in Fig.~(\ref{Fig1}). The Supplementary Material (see below) contains the results for six additional subjects. In all cases, the data were binned into integer distance values to decrease the statistical fluctuations (especially at long distances) without having to average across several subjects. 
Two different regimes are observed in the correlation function. At short distances, we see a collapse of the different
curves so Eq.~\eqref{EqScaling} holds indicating that the system is self-similar.
At longer distances, instead of the usual decay of the correlation function that is observed in physical models,
there is a plateau, probably due to the brain bilateral symmetry, as seen previously across cortical bilaterally
homologous regions by Salvador {\em et al.} \cite{salvador2,salvador},
and then an increase in the correlations, probably due to surface effects.

After one coarse-graining step, the length of a side of a block-voxel
is twice the size of a voxel, but then we need to renormalize the Euclidean distance in the coarse-grained
system by a factor $2$ to recover distance in voxels. If the system under consideration was infinite,
then $\langle C(r)\rangle$ would be defined for all $r$ irrespectively of the level
of description and they would be identical, see Eq. \eqref{EqScaling}.
However, as the system we consider is finite, there exists a maximum distance
$r_m$ in the original system. After one coarse-graining step, this maximal distance
equal to $r_m/2$ and after $n$ coarse-graining steps, this maximum distance equals $r_m/2^n$,
measured in units of voxels at the $n$th coarse-graining level.
If Eq. \eqref{EqScaling} holds, then it means that the system behaves
in the same manner at all scales, up to the maximum distance possible.  
Furthermore, it allows for a data-collapse of the correlation
functions by first re  the distance $r$ in voxel units to Euclidean distances,
$r \mapsto r 2^n a$ where $a$ is the voxel spacing in Euclidean distance (say mm)
in the original data and then renormalizing the correlation functions by the factor $2^{-\alpha n}$,
where $\alpha$ is a subject-dependent parameter.
It is easy to check that if the correlation function depends on distance
through a power law with exponent $\beta$, that is, $\langle C(r)\rangle\propto r^{-\beta}$,
then the exponent $\alpha$  in the renormalizing factor of the correlation function must be equal to $\beta$.
The relation $\alpha=\beta$ is fulfilled for this regime.

Scaling is only expected to apply in an intermediate regime where the sizes of the coarse-grained voxels
are big compared with the smallest length scale cut-off in the data, that is, the voxel scale set by the scanner,
and the largest distance covered by the data, that is, the size of the brain.
This is indeed observed in Fig. (\ref{Fig1}) where the data collapse is
best for the three curves corresponding to the bigger block voxels.  
\begin{figure}[h]
\vspace*{-1cm}
\begin{center}
\includegraphics[width=0.38\textwidth]{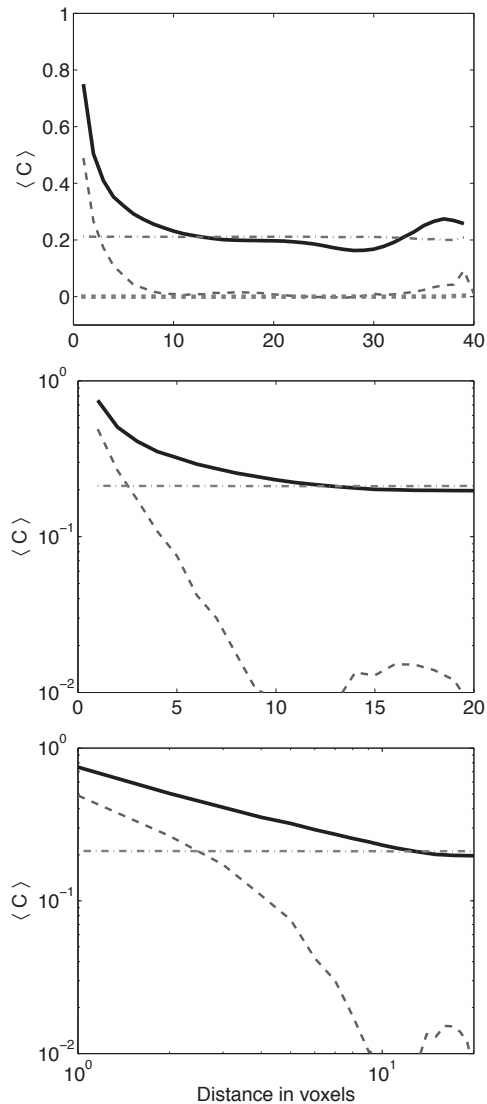}\\
 \end{center}
 \vspace*{-.3cm}
 \caption{The average correlation function $\langle C \rangle$ vs the voxel distance after one step
of coarse graining ($64\times 64\times 16$) for fMRI data (solid line)
spatially randomized voxels (dashed-dotted line) and temporal randomized voxels (dotted line).
Shown is also the average correlation function for phantom data (dashed line).
Top panel in linear-linear, middle in linear-log and bottom in log-log axis, respectively.
The correlation function for the fMRI data decays as a power law, while the
correlation function for the phantom data decays exponentially fast. The correlation function
for the spatially randomized data is constant while it is zero for the temporally randomized data.
}\label{Fig2}

\end{figure}

To verify that the shape of the average two-point correlation function is not due to
artifacts in the data, we randomized the positions of the voxels or the time series, see
Fig. (\ref{Fig2}).
By randomizing the positions, we expect to find a constant profile
for the correlations. In an infinite system or in a system where the correlation length is
much smaller than the system's size, this constant would be equal to zero, however, in
the system we study strong long-range correlations are present, hence the constant is
non-zero. By randomizing the time series of each voxel independently, we expect to destroy
all correlation among voxels and thus expect close to zero average correlation for all distances.
Furthermore, we also scanned a phantom to test our finding against artifact coming from the scanner.
The phantom data shows an exponential decay and no long range correlation,
except a peak at the longest distance that is also observed in brain data, confirming that
it is a surface effect.

\emph{Pondering the Correlation Function with the Voxels' Grey and White Matter Content:} When analyzing fMRI data, the focus is, unlike our approach, generally on grey matter voxels. Due to the large
millimeter size of the voxels, the discrimination between grey and white matter voxels
is based on a high resolution anatomical scan that gives the content of grey matter, white
matter and CSF for each voxels. A voxel is said to belong to grey/white matter if its content
in the latter exceeds a threshold that has to be fixed, the others are dismissed.
This procedure has the drawback that information is lost in the process, because the grey matter signal in dismissed voxels is lost.

Pure grey and white matter voxels are rare, but from
a high resolution scan, we know the content of grey and white matter of each
voxel. Hence, we can weight the correlations between two voxel
according to their relative content of grey and white matter with the following
normalization: $1=w^{g}+w^{w}+w^{CSF}$. We have performed the same analysis as we
did above. Using all voxels weighted with their contents of grey and white matter, respectively,
we calculated the average correlation functions
\begin{equation}
\langle C_{i,j}^{gg} \rangle =w_{i}^{g}w_{j}^{g} \langle C_{i,j} \rangle \quad \mbox{or} \quad \langle C_{i,j}^{\text{ww}} \rangle =w_{i}^{w}w_{j}^{w} \langle C_{i,j} \rangle.
\end{equation}
\begin{figure}[h]
\vspace*{-0.8cm}
\begin{center}
\hspace*{.1cm}
\includegraphics[width=0.38\textwidth]{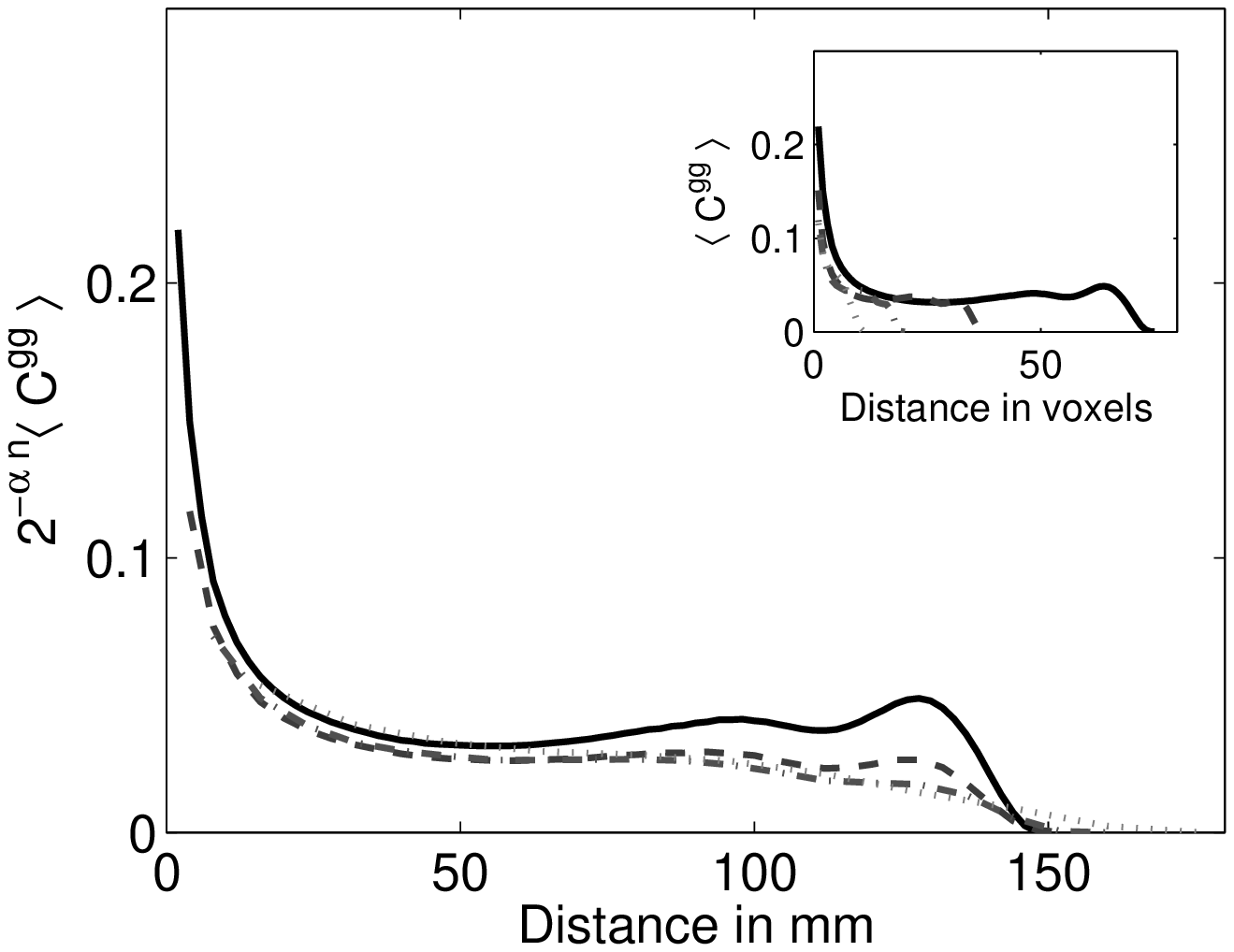}
\end{center}
\vspace*{-0.8cm}
\begin{center}
\includegraphics[width=0.38\textwidth]{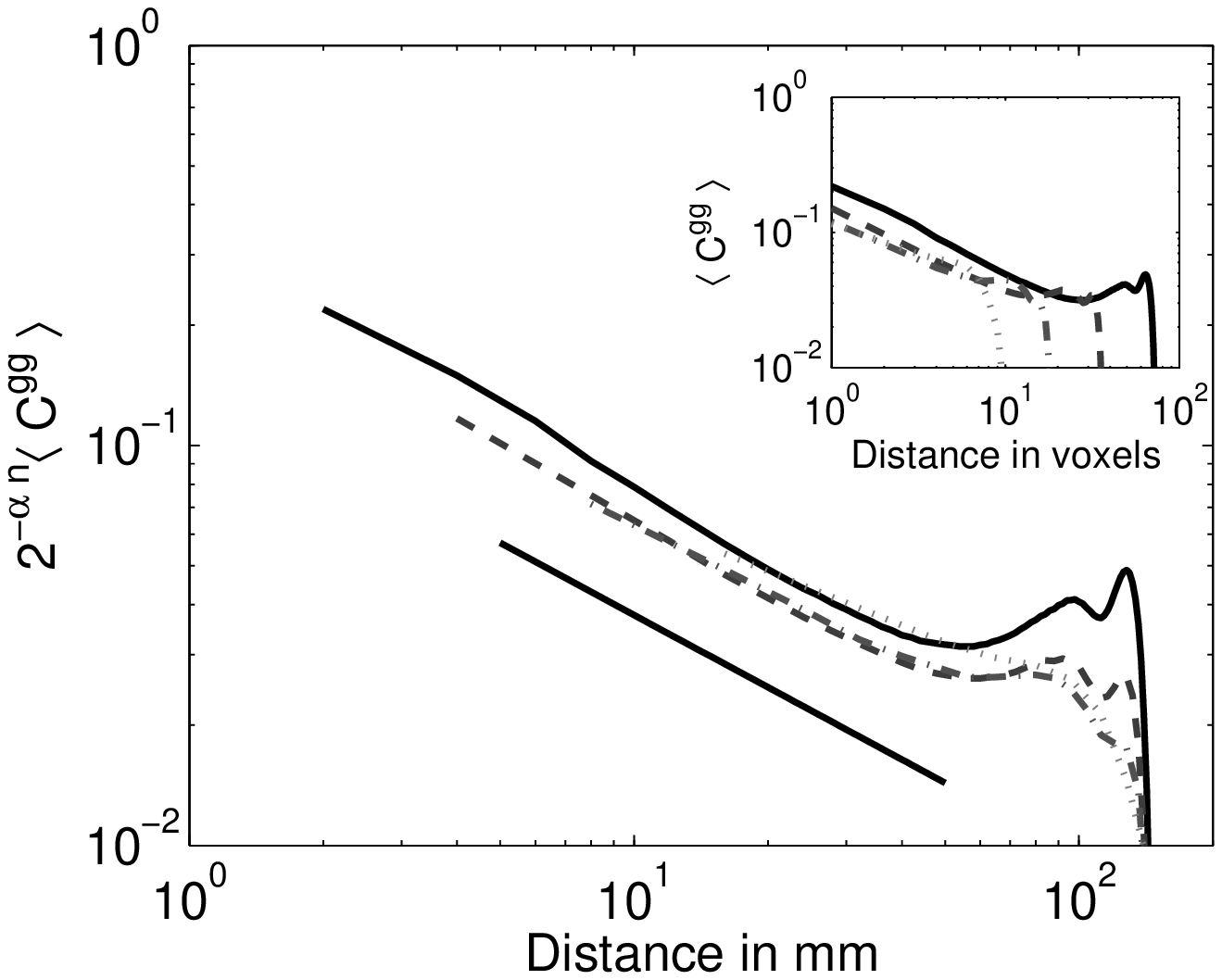}
\end{center}
\caption{The renormalized average correlation function $ 2^{-\alpha n} \langle C^{gg} \rangle$ of signal
weighted with the content of grey matter vs real distance
(inset: $\langle C^{gg} \rangle$ vs voxel distance) for the four levels of description for one subject.
Full line: $128\times 128\times 31$ ($n=0$), dashed line: $64\times 64\times 16$ ($n=1$), dashed-dotted
line: $32\times 32\times 8$ ($n=2$) and dotted line: $16\times 16\times 4$ ($n=3$).
Top panel linear-linear and bottom panel log-log axis, respectively.
Full straight line is a guide to the eye for a power law of exponent $\beta=0.60\pm0.2$.
Parameter $\alpha = 0.35$. }\label{Fig3}
\end{figure}
The correlation of grey matter is shown in Fig. \ref{Fig3}
and the correlation of white matter is shown in Fig. \ref{Fig4}.
The bump at the end of the distribution survives only in the grey matter,
which is to be expected, since it is probably due to a surface effect.
White matter correlations die off
smoothly before the surface of the brain is reached, which was expected since white matter is supposed to be enclosed in grey matter.
\begin{figure}[h]
\vspace*{-0.8cm}
\begin{center}
\includegraphics[width=0.38\textwidth]{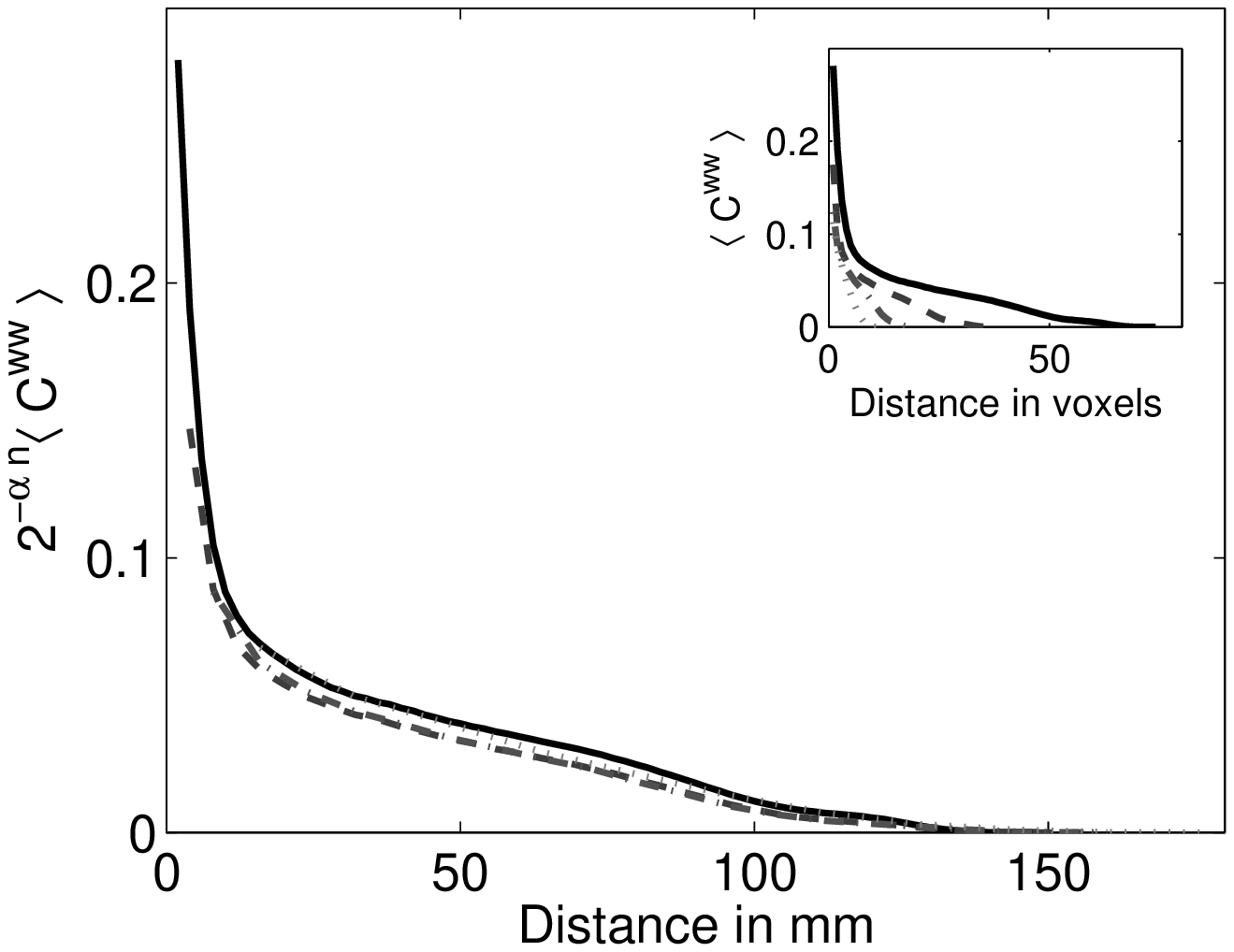}

\includegraphics[width=0.38\textwidth]{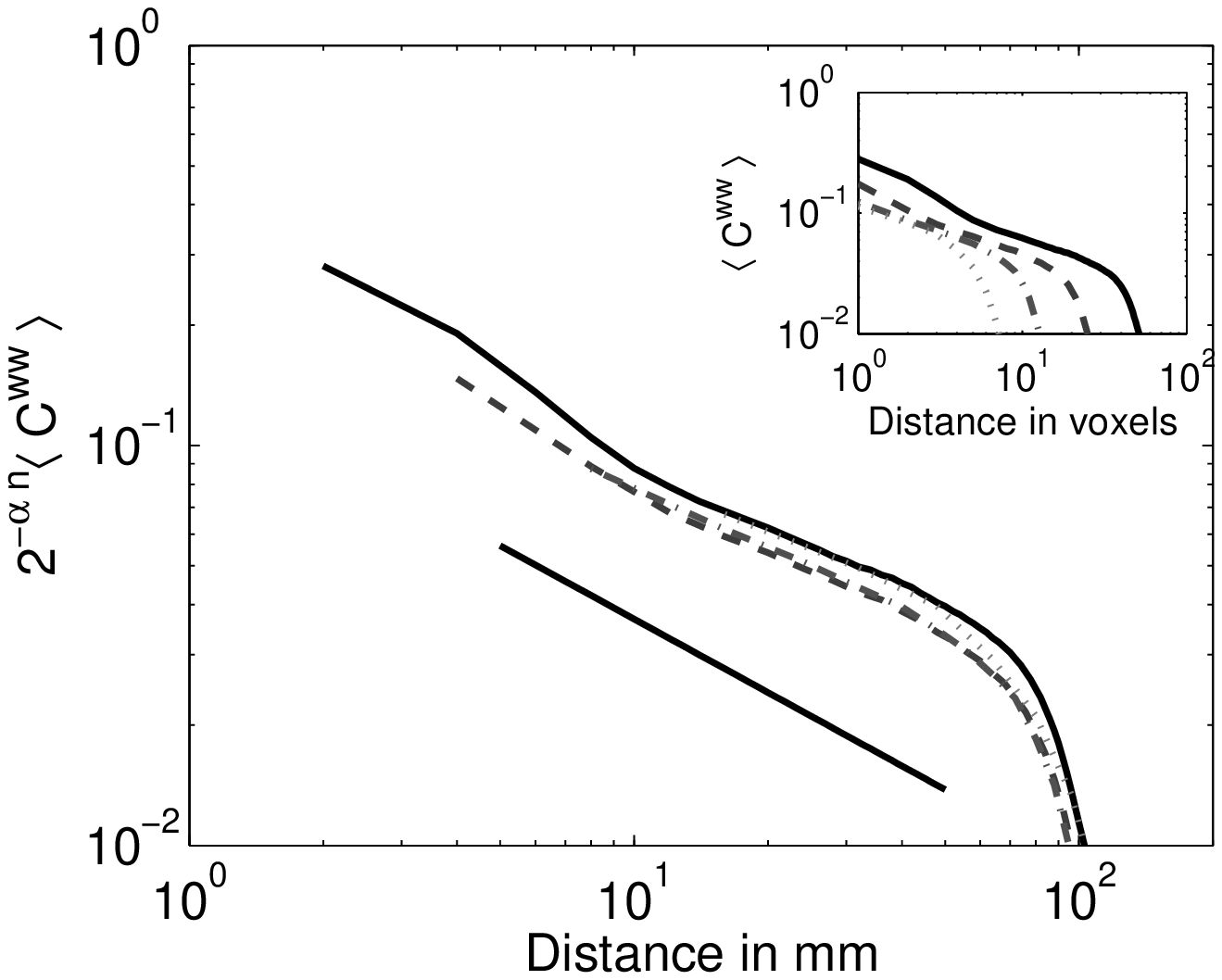}
\end{center}
\caption{The renormalized average correlation function $2^{-\alpha n} \langle C^{\text{ww}} \rangle$ of
signal weighted with the content of white matter vs real distance
(inset: $\langle C^{\text{ww}} \rangle$ vs voxel distance) for the four levels of description for one subject.
Full line: $128\times 128\times 31$ ($n=0$), dashed line: $64\times 64\times 16$ ($n=1$), dashed-dotted
line: $32\times 32\times 8$ ($n=2$) and dotted line: $16\times 16\times 4$ ($n=3$).
Top panel in linear-linear and bottom in log-log axis, respectively.
Full straight line is a guide to the eye for a power law of exponent $\beta=0.61\pm0.2$. Parameter $\alpha = 0.20$.}\label{Fig4}
\end{figure}
The fact that the exponents $\alpha\neq\beta$ in both Figs. \ref{Fig3} and \ref{Fig4} is related
to the deviation from power-law behavior of the correlation functions of the
grey and white matter. The scaling collapse is better
for the correlation function when all voxels are included (see Fig. \ref{Fig1})
than for the grey matter and white matter correlation functions
displayed in Figs. \ref{Fig3} and \ref{Fig4}.
It is remarkable that only when the entire fMRI BOLD signal (all voxels without thresholding) is included in the correlation matrix
we do obtain a good quality data collapse consistent with self-similarity. In general, we associate the essential dynamics and processing of the brain with grey matter, so our findings indicate that the separation of the fMRI BOLD signal using the weights $w^g$ and $w^w$ is more subtle than expected. 

\begin{figure}[h]
\begin{center}
\includegraphics[width=0.38\textwidth]{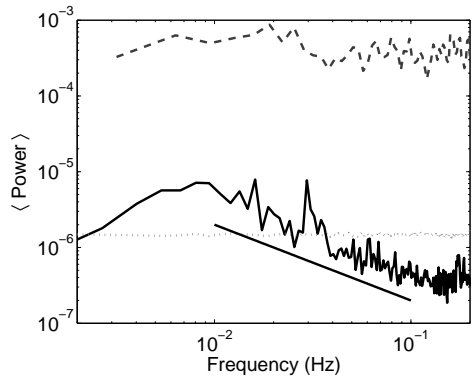}
\end{center}
\caption{Average power spectrum for the fMRI time series (solid line), the phantom time series (dashed line) and the
randomized time series (dotted line). Full straight line is a guide to the eye for a $1/f$ spectrum.}\label{Fig5}
\end{figure}

\emph{Temporal Self-similarity:}
To complement the spatial correlation analysis, we present the average power spectra of the fMRI time series. Temporal auto-correlations and power spectra have been studied by use of fMRI for a while. A comprehensive review is given in \cite{Woolrich-2001} and the use of these as a diagnostic tool was discussed in \cite{Bullmore-2005}. To obtain the spectra, we computed the spectrum of each voxel individually and normalized its integral to $1$.
This corresponds to analyzing the Fourier transform of the {\em auto-correlation} function.
We note that the term ``power spectrum'' is often used to denote the Fourier transform of the auto-covariance function.
However, it is better to operate with the normalized correlation function to avoid signals with more
power to totally dominate the average power spectra.
Although the frequency range we can explore is rather small, the power spectrum has a
reasonable $1/f$ dependence, thus showing criticality in time as well as in space, see Fig. (\ref{Fig5}).
To check against potential  artifacts of the measurement, we calculated the power spectra of phantom data and surrogate data, constructed by randomizing the brain fMRI time series. In both cases, we obtain a flat spectra as shown in Fig. (\ref{Fig5}) rejecting the possibility that the spectral features arise from scanning artifacts.
The peak at $0.03\mbox{Hz}$ superimposed on the $1/f$ slope corresponds to the low-frequency fluctuations dynamics described already for the so-called brain resting-state networks \cite{beckmann2005}.
\section{Summary and Discussion}
These results are the first direct demonstration of spatial and temporal self-similarity of the
brain resting state dynamics. Although several reports already hinted about the possibility that at large-scale the brain may be operating in such state, previous measures have been indirect and/or model dependent.
In contrast, in the present work scale invariance in space and time is determined from the entire 
data set without discarding any voxels and using the same renormalization techniques 
championed in the study of critical phenomena in physical systems and models. 

It is interesting to relate our finding of scale invariance with the  spatial and temporal localized
functional activity extracted from fMRI data by Independent Component Analysis (ICA) methods.
By construction, ICA emphasize the identification of strongly independent components and ``filtering'' out
less important contributions to the total BOLD signal.
Recent studies have investigated the growing complexity of the spatial structures
extracted by ICA as the total numbers of components is allowed
to increase \cite{Beckmann-2009, Kiviniemi-2009}. The additional structure obtained
this way is indeed consistent with the self-similar character
we have identified in the correlation function of the BOLD signal. 

It should be mentioned  that the distribution of BOLD fMRI spatiotemporal correlations was found already to be altered in certain chronic diseases \cite{baliki} which together with the present result seem to suggest that scale invariance could be an important objective marker of brain well-being both in health and disease. 

From a dynamical systems perspective, the uncovered self-similarity implies that the brain dynamics is permanently at an intermediate state between two extremes, one that is strongly correlated across large distances, producing transient highly integrated cortex states and the other in which only nearby clusters are acting in sync.  This scenario, of long range correlations in space and time, is only conceivable in dynamical systems at criticality and could be the manner in which the cortex can manage to produce an arbitrarily large repertoire of interaction patterns among arbitrarily distant cortical sites.

\section{Materials and Methods}
\emph{Image Acquisition:}
A 3T Philips MRI scanner was used to acquire T$_2^*$-weighted echo-planer images (EPI)
in $128 \times 128 \times 31$ voxels of dimension
$2.5 \times 2.5 \times 5.0\,\mbox{mm}^3$ using a repetition time of
$2000\, \mbox{ms}$, echo time of $60\, \mbox{ms}$ and a flip angle $90^\circ$.
An 8-channel array coil and SENSE factor $2$ were used as well as second-order shims.
Finally, a $3$-dimensional mask was used to identify the content of white matter, gray matter and CSF.

\emph{Image Pre-processing:}
The brain data consist of fMRI data sets of seven young adults healthy subjects in the resting state. 
The subjects were instructed to lie still in the scanner with their eyes closed avoiding falling asleep.  A total of $305$ functional volumes of each subject were acquired from each session, the first five of which were discarded in order to remove the effect of T$_1$ equilibration. Image pre-processing was performed using the University
of Oxford's FMRIB software library (FSL) involving realigning to account for movement using  FMRIB's Linear Image
Registration Tool (MCFLIRT) \cite{Jenkinson-2002}, high-pass temporal filtering using FEAT to remove low frequency artifacts and a high pass filter cutoff preset to $100\, \mbox{s}$.

\emph{Coarse Graining of the Signal:} Coarse graining is a well established technique 
to describe system's changes at different levels of spatial or temporal observation. 
This allows one to investigate to what extent the spatial or
temporal structures of a given phenomena exhibits scale invariance, see for example Ref. \cite{Binney-1992}.
A self-similar or scale invariance object, for instance a cauliflower, looks like itself 
at all scales. Each little bouquet of the cauliflower looks a
miniature version of the entire cauliflower. Hence, a short length scale study of the
cauliflower, in which one probes the substructure of an individual bouquet, will provide the same information as
the study at a large length scale where one observe the cauliflower as a composition of bouquets. 

\begin{figure}[h]
\begin{center}
 \includegraphics[width=0.48\textwidth]{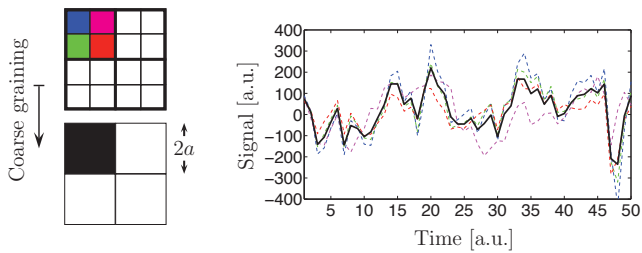}
 \end{center}
\caption{(Left) Example of coarse graining in 2 dimensions where there are 4 boxes ${\cal B}$ within a block-box ${\cal B'}$.
(Right) The four dashed-colored signals from the four original boxes ${\cal B}$ are averaged to
produce the solid-black coarse-grained signal of ${\cal B'}$.}\label{Fig6}
\end{figure}

In general, the approach to probe the behavior at different length scales consists 
in replacing the signal at a given point by a spatial average over a region of a given spatial extend about this point (see Fig.~\ref{Fig6}).
In the case of an fMRI signal, the procedure consists in aggregating $2 \times 2 \times 2 = 8$ adjacent voxels to create a block-voxel ${\cal B}'$ whose
BOLD intensity at time $t$, $v_{{\cal B}'}(t)$, is the average intensity of the BOLD intensities
at time $t$, $v_{\cal B}(t)$ in its constituting voxels ${\cal B} \in { \cal B}'$, that is,
\begin{equation}
   v_{{\cal B}'}(t) = \frac{1}{8} \sum_{{\cal B} \in {{\cal B}'}}v_{\cal B}(t),
\label{block}
\end{equation}
see Fig.~\ref{Fig6} for an illustration.

Computing Eq.~\ref{block} for all times $t = 1,2,\ldots,300$ gives the time series for the BOLD signal in the
block-voxels $\left\{v_{{\cal B}'}(t)\right\}$. 
We now repeat the coarse-graining procedure on the time series for the BOLD signal in the
block-voxels $\left\{v_{{\cal B}'}(t)\right\}$ to produce a time series for the BOLD signal in the
block-voxels $\left\{v_{\cal B''}(t)\right\}$ and so on. For the spatial resolution of our fMRI data, 
the coarse-graining procedure applied three times yield voxels ${\cal B} (128 \!\times\! 128 \!\times\! 31) \mapsto {\cal B'} (64 \!\times\! 64 \!\times\! 16)
\mapsto  {\cal B''} (32 \!\times\! 32 \!\times\! 8) \mapsto  {\cal B'''} (16 \!\times\! 16 \!\times\! 4)$
with associated signals $v_{\cal B}(t)  \mapsto  v_{\cal B'}(t)  \mapsto v_{\cal B''}(t) \mapsto  v_{\cal B'''}(t)$.
In the first coarse-graining step, the ultimate plane of voxels at the $31^{th}$ slice is coarse grained only along
the $x-y$ plane. Averages has been calculated over active voxels only to prevent the inclusion of non-active voxels at the boundaries. 

Next in our analysis, we consider the time series of each (block) voxel as a realization of a random variable and
compute the time correlation matrix at the four levels of coarse graining. This matrix is obtained by
computing the correlation between all pairs of (block) voxels ${\cal B}_i$ at position $\mathbf{r}_i$ and ${\cal B}_j$ at
position $\mathbf{r}_j$:
\begin{equation}
C_{\cal B}\left(v_{{\cal B}_i},v_{{\cal B}_j}\right)= \frac{ \frac{1}{T} \sum\limits_{t=1}^{T} v_{{\cal B}_i}(t)v_{{\cal B}_j}(t)-\hat{\mu}_{{\cal B}_i}\hat{\mu}_{{\cal B}_j}}
    { \hat{\sigma}_{{\cal B}_i}\hat{\sigma}_{{\cal B}_j}},
    \label{corr}
\end{equation}
where $\hat{\mu}_{{\cal B}_i}=\frac{1}{T}\sum_{t=1}^{T}v_{{\cal B}_i}(t)$ is the temporal average value of (block)
voxel's ${\cal B}_i$ BOLD signal, and
$\hat{\sigma}_{{\cal B}_i}=\sqrt{\frac{1}{T}\sum_{t=1}^{T} v^2_{{\cal B}_i}(t)-\hat{\mu}^2_{{{\cal B}_i}}}$
is the standard deviation of the BOLD signal of (block) voxel ${\cal B}_i$ and $T = 300$.

\emph{Scaling:} From the correlation matrix, the correlation function at
equal time $C_{\cal B}\left(v_{{\cal B}_i},v_{{\cal B}_j}\right)$ is calculated as a function of the position in space of the voxels.
We then compute the correlation function  $\langle C_{\cal B}(r) \rangle$ averaged over all voxels with
$r = \left|\mathbf{r}_i - \mathbf{r}_j\right|$ measured in units of (block) voxels.
This is done for each level of coarse graining to obtain $\langle C_{\cal B}(r) \rangle$, $\langle C_{\cal B'}(r) \rangle$,
$\langle C_{\cal B''}(r) \rangle$ and $\langle C_{\cal B'''}(r) \rangle$ 
and the indices ${\cal B}, {\cal B'}, {\cal B''}$ and ${\cal B'''}$ indicate the level of coarse graining

Let us discuss what to expect concerning the relation of the correlation function calculated
at one coarse-graining level, say $\langle C_{\cal B}(r) \rangle$, compared to the one at
the next level, $\langle C_{\cal B'}(r) \rangle$.
To clarify the effect of coarse graining, we relate the correlation function at this level of
coarse graining to the {\em covariance function} at the previous level. For a homogeneous system,
the covariance function is given by
\begin{equation}
\mbox{Cov}_{\cal B}(r)=\langle v_{\cal B}(0)v_{\cal B}(r)\rangle- \langle v_{\cal B}(0)\rangle\langle v_{\cal B}(0)\rangle,
\end{equation}
where ${\langle \cdot \rangle}$ denotes spatial and temporal average.
We consider, like in Eq. (\ref{block}), the coarse graining over a block-voxel ${\cal B'}$ centered
at ${\bf r}$:
\begin{equation}
v_{\cal B'}({\bf r})=\frac{1}{|{\cal B}|}\sum_{ {\bf r + a} \in {\cal B'} }v_{\cal B}({\bf r+a}),
\end{equation}
where the block-voxel ${\cal B'}$ contains $|{\cal B}|$ voxels.
We then have
\begin{equation}
\langle C_{\cal B'}(r) \rangle =\frac{ \sum_{{\bf a}_1,{\bf a}_2} \mbox{Cov}_{\cal B}({\bf r}+{\bf a}_2-{\bf a}_1)} {\sum_{{\bf a}_1,{\bf a}_2} \mbox{Cov}_{\cal B}({\bf a}_2-{\bf a}_1)}.
\label{coarse}
\end{equation}
For a scale invariant (self-similar) system of infinite size, $\langle C_{\cal B'}(r) \rangle$ will remain invariant if distance
$r$ is measured in units of coarse-graining boxes, that is,
\begin{equation}
\langle C_{\cal B}(r)\rangle = \langle C_{\cal B'}(r)\rangle = \langle C_{\cal B''}(r)\rangle =
\langle C_{\cal B'''}(r)\rangle
\label{EqScaling}.
\end{equation}
Note that $r$ is the distance expressed in the voxel unit corresponding to the given
level of coarse graining.

\begin{figure}[h]
\hspace*{1.1cm}\includegraphics[width=0.30\textwidth]{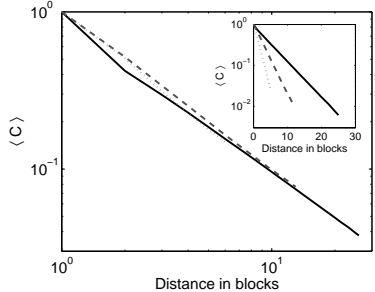}
\caption{Main: The average correlation function $\langle C \rangle$ vs the block distance
for three coarse-graining levels of an hypothetical object with a power-law form of the correlation function.
The graphs collapse in voxel distance. Inset: Results from an hypothetical object having an exponential
form of the correlation function where the curves do not collapse in block distance.}\label{Fig7}
\end{figure}
The simplest case of scale invariance corresponds to a power-law behavior of $\langle C(r) \rangle$. 
Figure~\ref{Fig7} illustrates the different behavior under rescaling of two 3-dimensional artificial systems, 
where we imposed the correlation to have either a power-law or an exponential behavior:
\begin{subequations}
\begin{align}
    \langle C(r) \rangle &=\frac{1}{1+r}, \label{EqScaling2} \\
    \langle C(r) \rangle &=\exp(-r/5).
\end{align}
\end{subequations}
Note that only the correlations having a power-law form
do collapse in block unit. 

\begin{acknowledgments}
We thank Professor Joseph Hajnal for providing the phantom data. It is a great pleasure to acknowledge very helpful discussion with Dr. Christian F. Beckmann. The project was supported by the EPSRC under  grant EP/E049451/1. DRC is supported by NINDS (USA).
\end{acknowledgments}

\setcounter{figure}{0}
\renewcommand{\thefigure}{S\arabic{figure}}
\begin{widetext}
\newpage

\begin{center}{\large \bf Supplementary information}\end{center}

The three average correlations $\langle C\rangle$, $\langle C^{gg}\rangle$ and $\langle C^{ww}\rangle$ at four levels
of description for six additional subjects are consistent with the those presented in the main text. 
In general we find that the equality $\alpha=\beta$ is more accurately satisfied when only the three most coarse-grained data sets are included in the data collapse.  
Full line: $128\times 128\times 31$ ($n=0$), dashed line: $64\times 64\times 16$ ($n=1$), dashed-dotted
line: $32\times 32\times 8$ ($n=2$) and dotted line: $16\times 16\times 4$ ($n=3$).
Full straight line are guides to the eye for a power law with exponent $\beta$.

\begin{figure}[h]
\begin{center}
    \includegraphics[width=0.850\textwidth]{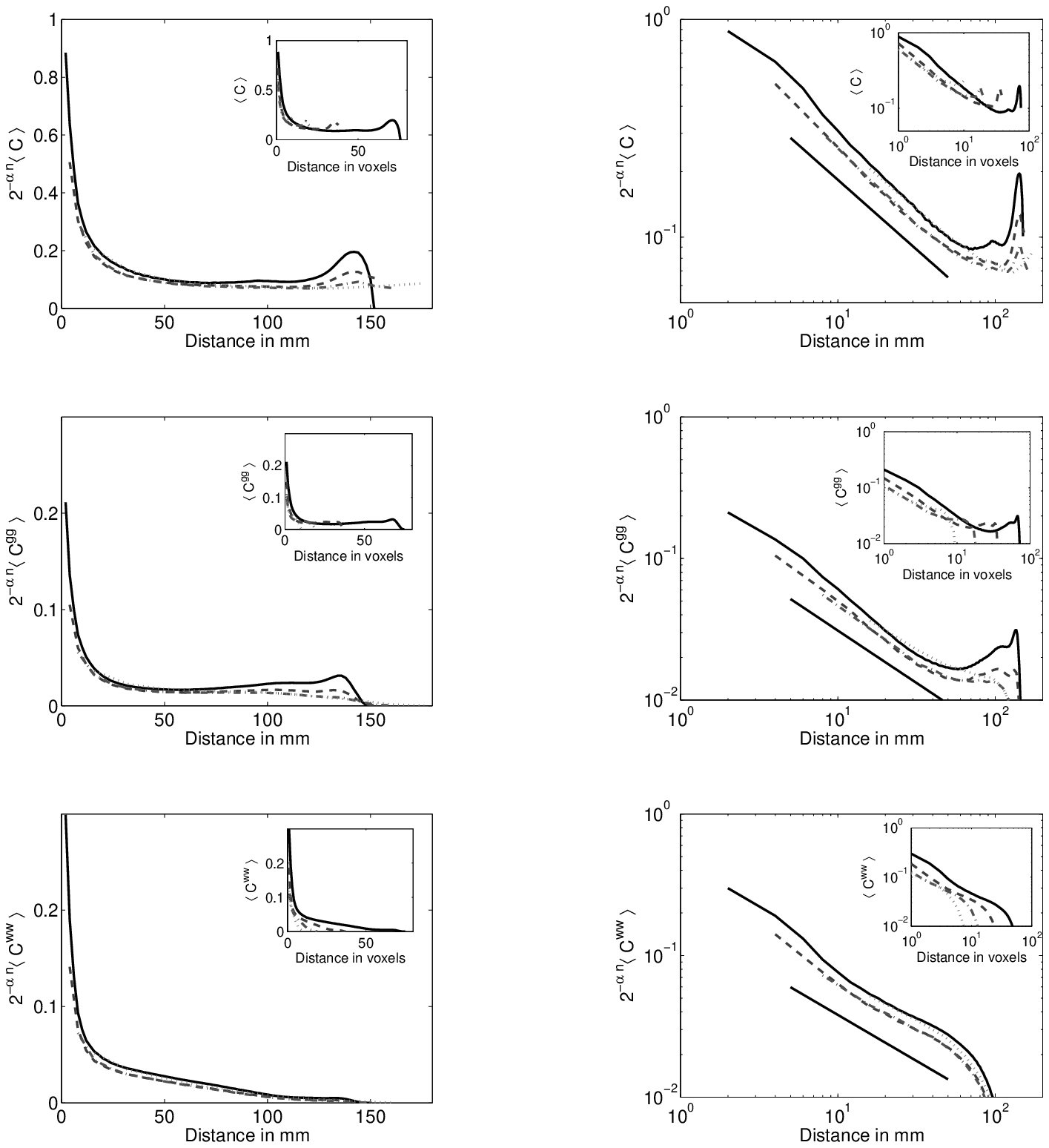}
\end{center}
\caption{The renormalized average correlation functions $2^{-\alpha n} \langle C \rangle$ vs real distance
(inset: $\langle C \rangle$ vs voxel distance). Left panel linear-linear and right panel log-log axis,
respectively. Top panel $2^{-\alpha n} \langle C\rangle$ with $\beta=0.64\pm 0.2$ and $\alpha=0.51$, middle panel
$2^{-\alpha n} \langle C^{gg}\rangle$ with $\beta=0.74\pm 0.2$ and  $\alpha=0.50$  and bottom panel $2^{-\alpha n} \langle C^{ww}\rangle$ with
$\beta=0.65\pm 0.2$ and $\alpha=0.40$, respectively.}\label{SI}
\end{figure}

\newpage

\begin{figure}[h]
\begin{center}
    \includegraphics[width=0.850\textwidth]{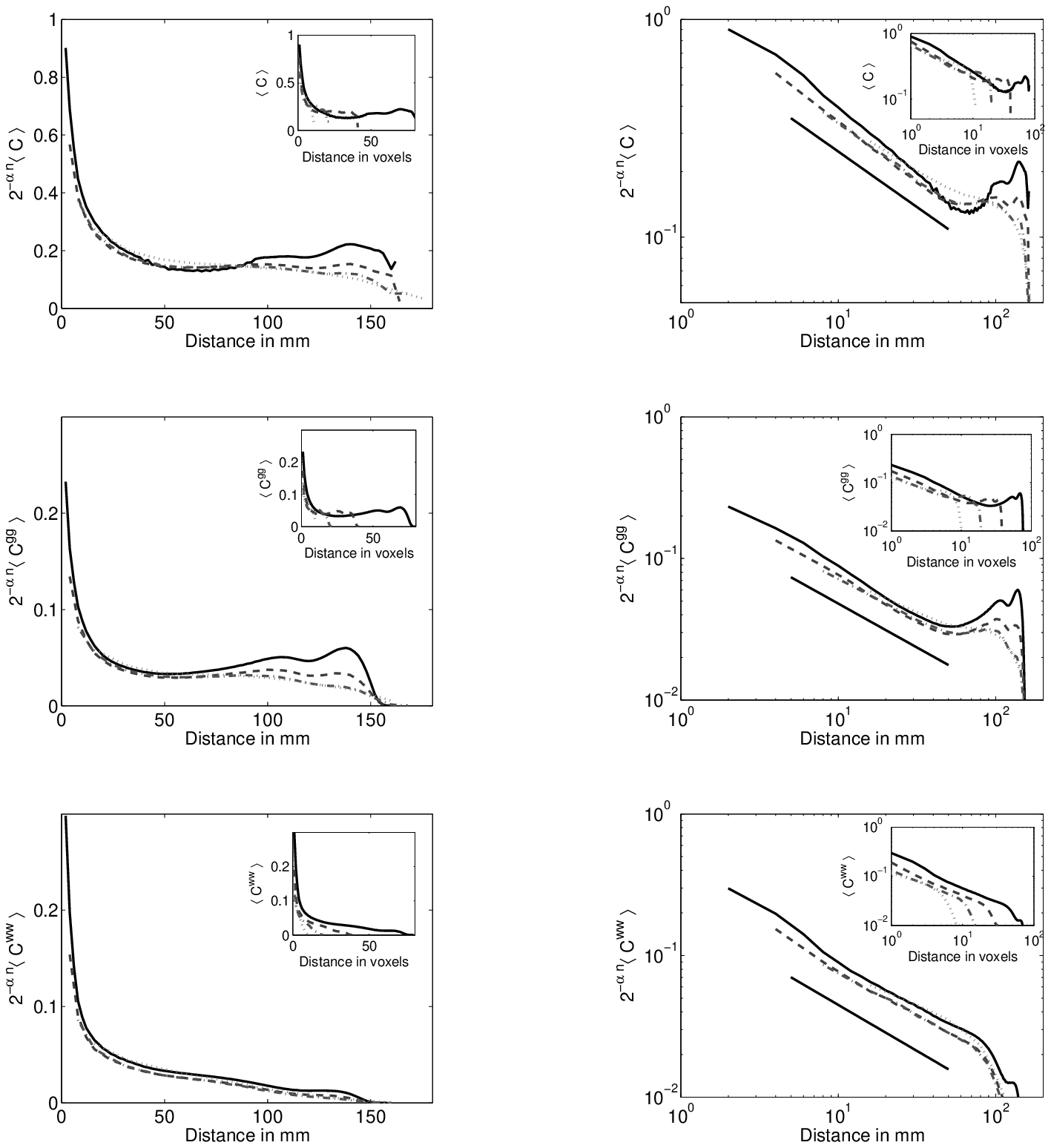}
\end{center}
\caption{The renormalized average correlation functions $2^{-\alpha n} \langle C \rangle$ vs real distance
(inset: $\langle C \rangle$ vs voxel distance). Left panel linear-linear and right panel log-log axis,
respectively. Top panel $2^{-\alpha n} \langle C\rangle$ with $\beta=0.51\pm 0.2$ and $\alpha=0.42$, middle panel
$2^{-\alpha n} \langle C^{gg}\rangle$ with $\beta=0.62\pm 0.2$ and  $\alpha=0.37$  and bottom panel $2^{-\alpha n} \langle C^{ww}\rangle$ with
$\beta=0.65\pm 0.2$ and $\alpha=0.31$, respectively.}\label{SI}
\end{figure}

\newpage

\begin{figure}[h]
\begin{center}
    \includegraphics[width=0.850\textwidth]{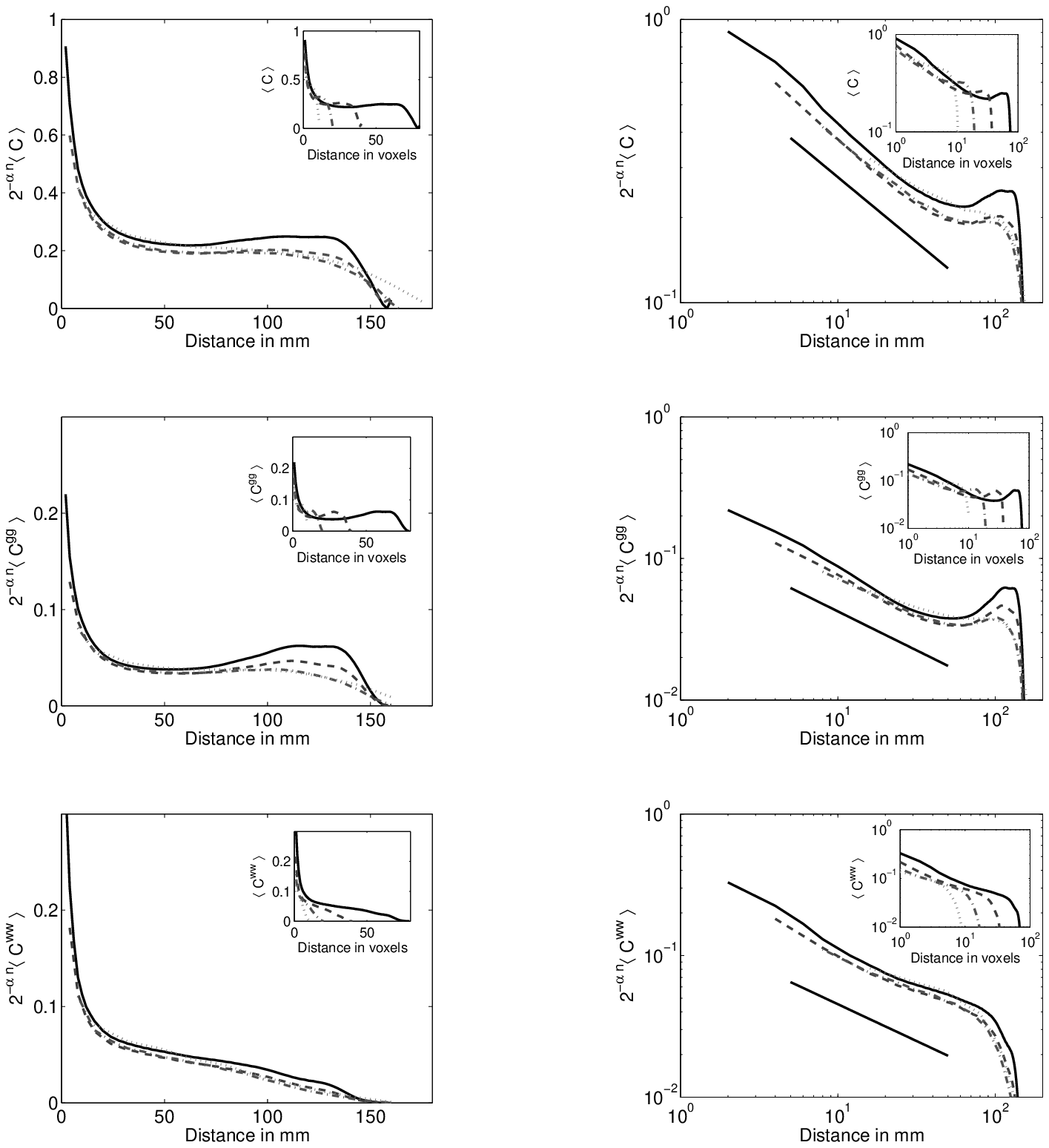}
\end{center}
\caption{The renormalized average correlation functions $2^{-\alpha n} \langle C \rangle$ vs real distance
(inset: $\langle C \rangle$ vs voxel distance). Left panel linear-linear and right panel log-log axis,
respectively. Top panel $2^{-\alpha n} \langle C\rangle$ with $\beta=0.46\pm 0.2$ and $\alpha=0.37$, middle panel
$2^{-\alpha n} \langle C^{gg}\rangle$ with $\beta=0.55\pm 0.2$ and  $\alpha=0.40$  and bottom panel $2^{-\alpha n} \langle C^{ww}\rangle$ with
$\beta=0.52\pm 0.2$ and $\alpha=0.25$, respectively.}\label{SI}
\end{figure}

\newpage

\begin{figure}[h]
\begin{center}
    \includegraphics[width=0.850\textwidth]{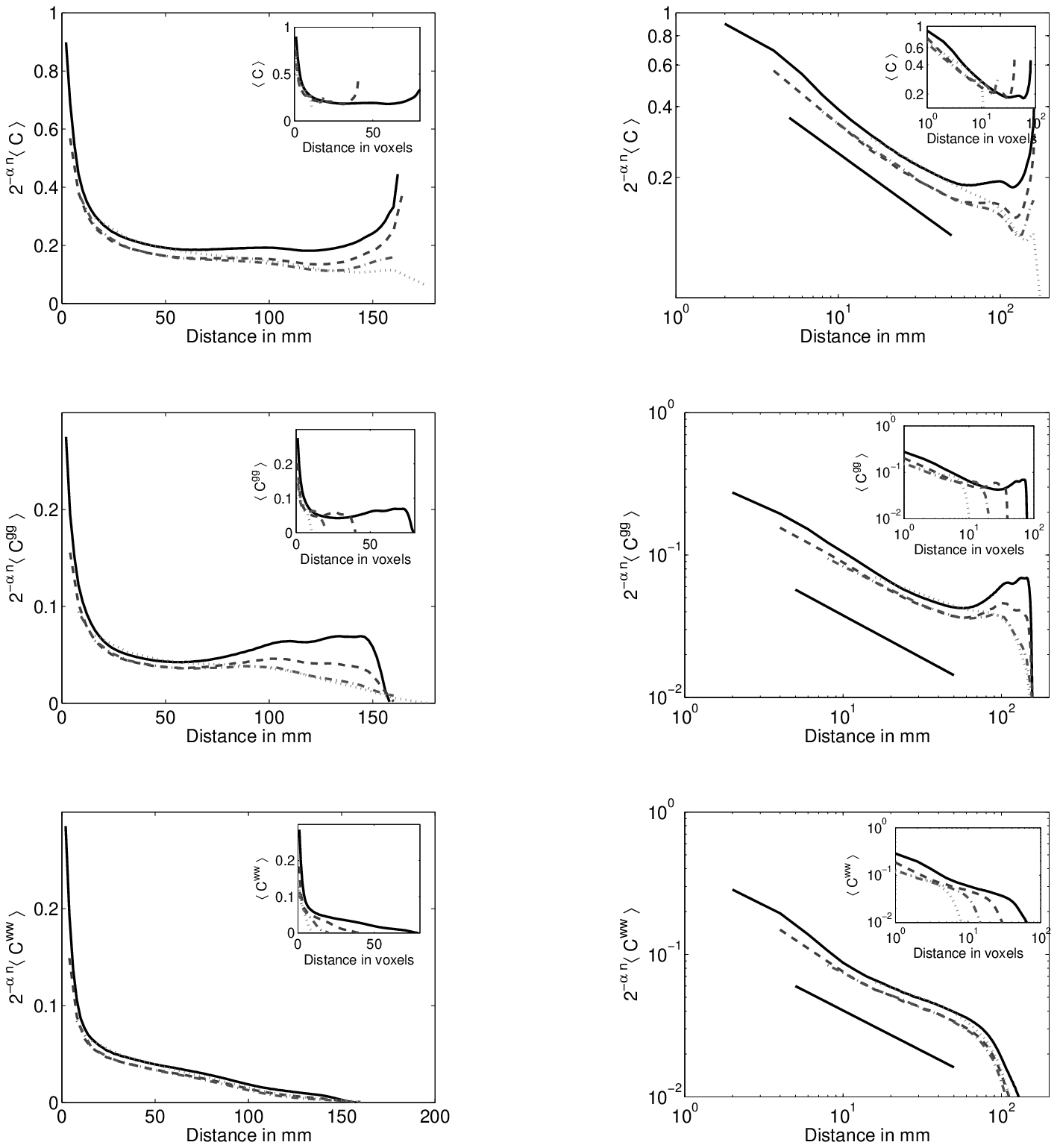}
\end{center}
\caption{The renormalized average correlation functions $2^{-\alpha n} \langle C \rangle$ vs real distance
(inset: $\langle C \rangle$ vs voxel distance). Left panel linear-linear and right panel log-log axis,
respectively. Top panel $2^{-\alpha n} \langle C\rangle$ with $\beta=0.50\pm 0.2$ and $\alpha=0.41$, middle panel
$2^{-\alpha n} \langle C^{gg}\rangle$ with $\beta=0.60\pm 0.2$ and  $\alpha=0.37$  and bottom panel $2^{-\alpha n} \langle C^{ww}\rangle$ with
$\beta=0.57\pm 0.2$ and $\alpha=0.30$, respectively.}\label{SI}
\end{figure}

\newpage

\begin{figure}[h]
\begin{center}
    \includegraphics[width=0.850\textwidth]{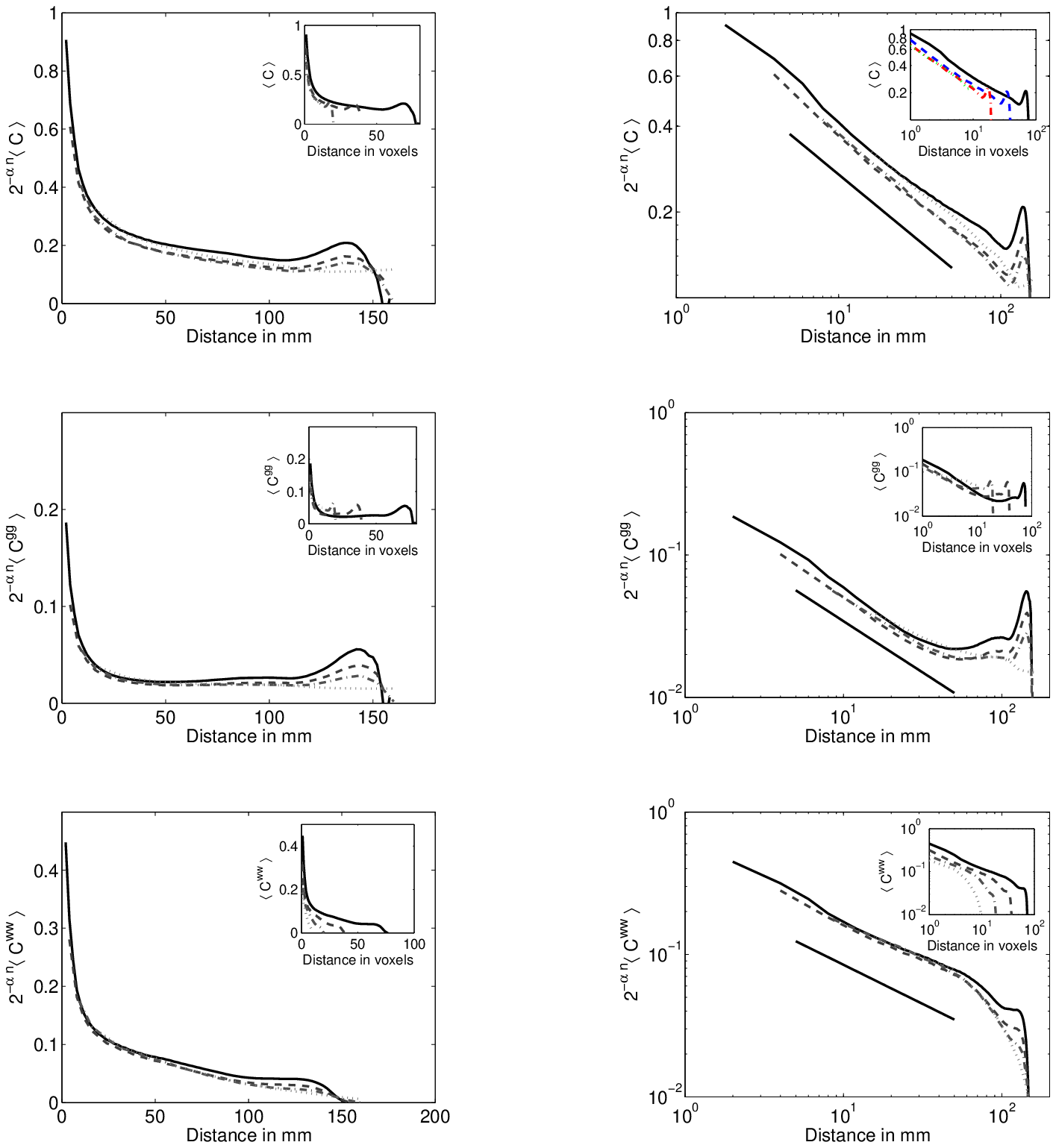}
\end{center}
\caption{The renormalized average correlation functions $2^{-\alpha n} \langle C \rangle$ vs real distance
(inset: $\langle C \rangle$ vs voxel distance). Left panel linear-linear and right panel log-log axis,
respectively. Top panel $2^{-\alpha n} \langle C\rangle$ with $\beta=0.47\pm 0.2$ and $\alpha=0.34$, middle panel
$2^{-\alpha n} \langle C^{gg}\rangle$ with $\beta=0.72\pm 0.2$ and  $\alpha=0.58$  and bottom panel $2^{-\alpha n} \langle C^{ww}\rangle$ with
$\beta=0.55\pm 0.2$ and $\alpha=0.17$, respectively.}\label{SI}
\end{figure}

\newpage

\begin{figure}[h]
\begin{center}
    \includegraphics[width=0.850\textwidth]{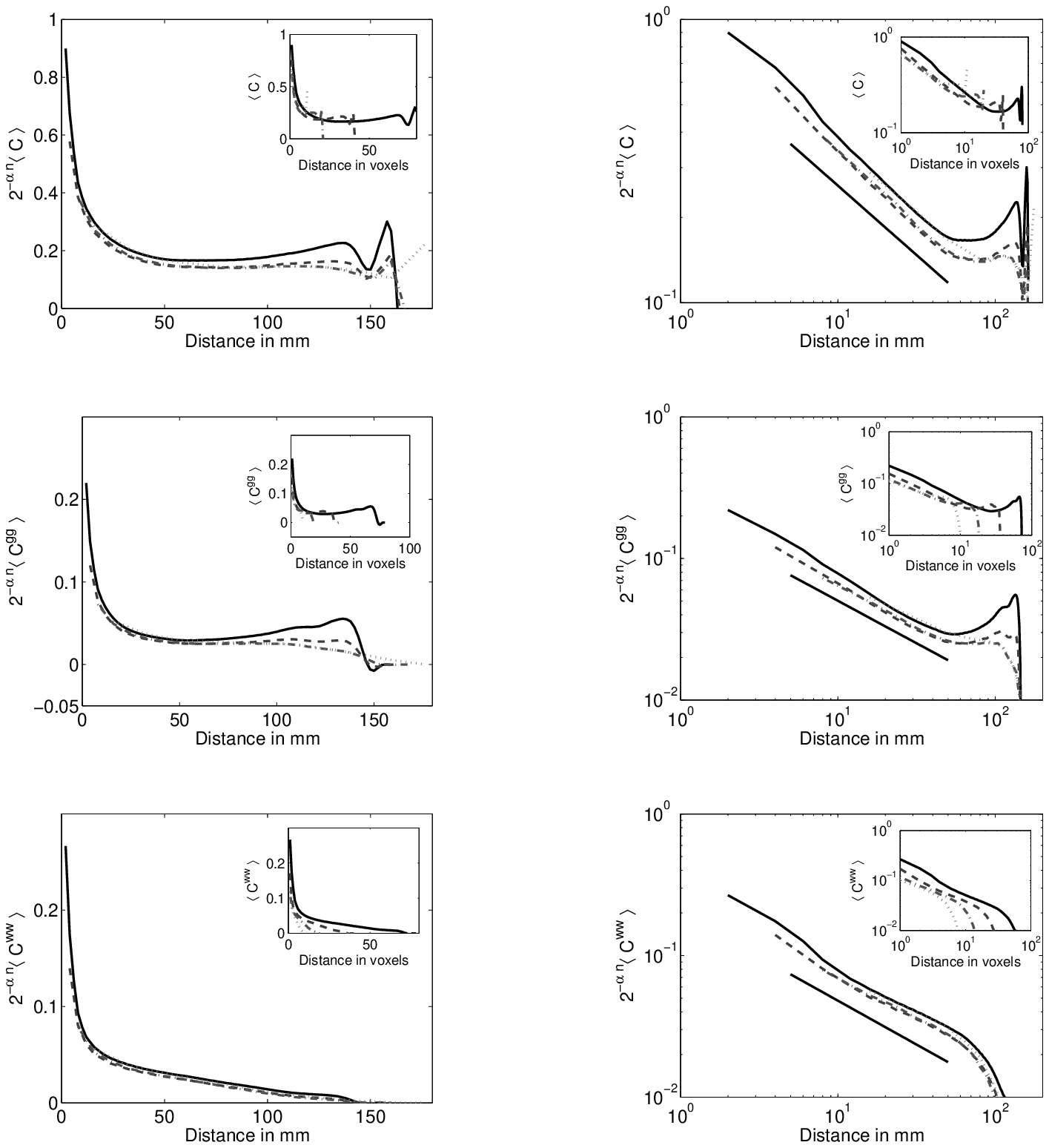}
\end{center}
\caption{The renormalized average correlation functions $2^{-\alpha n} \langle C \rangle$ vs real distance
(inset: $\langle C \rangle$ vs voxel distance). Left panel linear-linear and right panel log-log axis,
respectively. Top panel $2^{-\alpha n} \langle C\rangle$ with $\beta=0.49\pm 0.2$ and $\alpha=0.39$, middle panel
$2^{-\alpha n} \langle C^{gg}\rangle$ with $\beta=0.60\pm 0.2$ and  $\alpha=0.36$  and bottom panel $2^{-\alpha n} \langle C^{ww}\rangle$ with
$\beta=0.62\pm 0.2$ and $\alpha=0.29$, respectively.}\label{SI}
\end{figure}

\end{widetext} 

\end{document}